\documentstyle[onecolumn,astrobib,psfig]{mn-ab}

\title[Structure in the Ly-$\alpha$ forest]
{Large-scale structure in the Lyman-$\alpha$ forest -- A new technique}

\author[J. Liske et al.]{J.~Liske,$^1$ J.~K.~Webb$^1$ and R.~F.~Carswell$^2$\\
	$^1$School of Physics, University of New South Wales, Sydney 2052,
	Australia.\\
	$^2$Institute of Astronomy, Madingley Road, Cambridge CB3 0HA, U.K.}

\date{Accepted
...... Received .....}

\pagerange{\pageref{firstpage}--\pageref{lastpage}}

\pubyear{1998}

\newcommand{\lya}{\mbox{Ly$\alpha$}}
\newcommand{\be}{\begin{equation}}
\newcommand{\ee}{\end{equation}}
\newcommand{\bea}{\begin{eqnarray}}
\newcommand{\eea}{\end{eqnarray}}
\newcommand{\mref}[1]{(\ref{#1})}

\begin{document}

\label{firstpage}
\maketitle

\begin{abstract}
We present a new technique for detecting structure on Mpc scales in
the \lya\ forest. The technique is easy to apply in practice since it
does not involve absorption line fitting but is rather based on the
statistics of the transmitted flux. It identifies and assesses the
statistical significance of regions of over- or underdense \lya\
absorption and is fairly insensitive to the quality of the data. Using
extensive simulations we demonstrate that the new method is
significantly more sensitive to the detection of large-scale structure
in the \lya\ forest than a traditional two-point correlation function
analysis of fitted absorption lines.
\end{abstract}

\begin{keywords}
large-scale structure of Universe -- quasars: absorption lines
\end{keywords}

\section{Introduction}
Over the past few years new information has emerged that warrants a
new investigation into the large-scale clustering properties of the
\lya\ forest seen in the spectra of distant QSOs. Observationally,
major advances have been achieved with the help of the HST and the
Keck telescope. At the lowest redshifts, where \lya\ absorbers are
probed by HST, multi-slit spectroscopy of galaxies in the fields of
bright QSOs has resulted in the direct identification of galaxies
which produce \lya\ absorption, as evidenced by the anti-correlation
between the \lya\ equivalent width and the distance of the absorbing
galaxy from the QSO sight-line (\citeNP{Chen98}; \citeNP{Lanzetta95}).
The correlation appears to extend out to very large distances
\cite{Tripp98}, where the interpretation may be different and the
\lya\ absorber presumably may not be directly associated with the
`identified' galaxy. Regardless of the interpretation on any scale, it
now seems clear that the number density of \lya\ absorbers is larger
in those regions of space where galaxies reside, and thus \lya\
absorbers trace large-scale structure.

At high redshift, there is also mounting evidence that a significant
fraction of \lya\ absorbers is associated with (proto-) galaxies. The
discovery of C{\footnotesize IV} in 75 per cent of absorbers with
$N($H{\footnotesize I}$) = 3.0 \times 10^{14}$~cm$^{-2}$ at $z \sim 3$
\cite{Songaila96} challenges the original interpretation of the \lya\
forest as a primordial, randomly distributed, intergalactic
population. Furthermore, \citeN{Fernandez96} showed that \lya\ lines
with associated weak C{\footnotesize IV} absorption cluster strongly
in redshift and they concluded that the observed clustering is broadly
consistent with that expected for galaxies at $z \sim 2 - 3$ (but see
also \citeNP{Songaila96}).

The high redshift observations may be understood theoretically and
placed within the context of cosmological structure formation with the
help of numerical simulations. Using a uniform metal enrichment of the
IGM of [C/H] $\sim -2.5$, produced by a postulated Population III
burst of star formation, \citeN{Hellsten97} found that they could
reproduce the observed mean value of the C{\footnotesize
IV}/H{\footnotesize I} ratio with numerical simulations of
cosmological structure formation. However, the scatter of this ratio
implied an inhomogeneous metallicity distribution in the
IGM. \citeN{Gnedin98} subsequently suggested that the dominant
mechanism for the enrichment of the IGM is the merger mechanism which
reproduces both the mean and scatter of the C{\footnotesize
IV}/H{\footnotesize I} ratio.

In general, the simulations seem to suggest that \lya\ absorbers are a
less biased tracer of the underlying mass distribution than are
galaxies (\citeNP{Cen94}; \citeNP{Miralda96}; \citeNP{CenSim97};
\citeNP{Hernquist96}; \citeNP{Zhang95}; \citeNP{Petitjean95};
\citeNP{Muecket96}; \citeNP{Riediger98}; \citeNP{Wadsley97};
\citeNP{Bond98}).

There is also direct observational evidence that the \lya\ forest
exhibits large-scale structure. \citeN{Pando96} used a discrete
wavelet transform to perform a space-scale decomposition of the \lya\
forest and to demonstrate the existence and evolution of clusters on
scales as large as $20~h^{-1}$~Mpc. Recently, \citeN{Williger98}
reported correlations of \lya\ absorbers over $\sim 36~h^{-1}$
comoving Mpc in the plane of the sky at $2.15 < z < 3.37$.

Fitting individual absorption lines and computing their two-point
correlation function (tpcf) is the most commonly adopted approach to
clustering analysis of the \lya\ forest. \citeN{Pando96} discussed
this and other methods based on line statistics and concluded that a
space-scale decomposition is most effective. However, the analysis by
\citeN{Fernandez96} demonstrates the difficulty of using any sort of
analysis based on the statistics of fitted absorption lines. Even in
high resolution spectra blending successfully masks even very strong
clustering, so that any procedure involving identifying individual
absorption lines may severely underestimate the strength and scale of
the `true' correlation. In addition, if the aforementioned numerical
simulations are more or less correct then at least the low column
density forest does not correspond to well-defined individual `clouds'
since it arises in a fluctuating but continuous medium with small to
moderate overdensities.

Ideally we therefore need a statistical method which does not rely on
identifying individual lines, and which is free from any systematic
effects associated with line counting. In this paper we introduce a
new technique based on the statistical properties of the transmitted
flux. The method is a space-scale decomposition and as such retains
spatial information. It allows us to locate specific structures in the
\lya\ forest, and assess their significance, as compared to a random
distribution. The method is compared to a line counting/tpcf method,
and we show that it is substantially more sensitive.

The organisation of this paper is as follows: in section
\ref{technique} we describe the new analysis and carry out all
necessary analytic calculations. In section \ref{simulations} we use
Monte-Carlo simulations to compare the new method with a tpcf
analysis. We present our conclusions in section \ref{conclusions}.

\section{Technique} \label{technique}
We base our analysis on the null-hypothesis that any \lya\ forest
spectrum can be fairly well represented by a collection of individual
absorption lines (\citeNP{Carswell84}; \citeNP{Kirkman97};
\citeNP{Lu96}; \citeNP{Hu95}) whose parameters are
uncorrelated. Usually those lines are taken to be Voigt profiles and
we shall adopt this although the exact shape of the profile is not
relevant. We also need to adopt the functional form of the
distribution of the absorption line parameters, $\eta(z, N, b)$, which
we take from observations. We stress that we make {\em no} assumptions
about what causes the absorption lines. Our analysis does not rely on
identifying an absorption line with an individual, well-defined
absorbing cloud. The composition of a spectrum of individual lines is
purely descriptive. We simply use the null-hypothesis to predict
integral properties of the absorption caused by the collection of
lines.

The general idea of the new analysis then is to use those predictions
to identify over- and underdense regions of absorption as a function
of scale and position (space-scale decomposition) and to assess their
statistical significance. This is implemented by using a matched
filter technique; in order to obtain an estimate of the mean
transmission we simply convolve a normalised spectrum (of $N_{\rm p}$
pixels) with a smoothing function of scale $\sigma_{\rm s}$ and repeat
this process for all possible scales ($\sigma_{\rm s} =
1,\ldots,N_{\rm p}$). When plotted in the ($\lambda, \sigma_{\rm s}$)
plane, this procedure results in the `transmission triangle' of the
spectrum. When using a top hat function as the smoothing function the
base of the transmission triangle is the spectrum itself (the original
spectrum smoothed by a top hat of width $\sigma_{\rm s} = 1$~pixel)
and the tip of the triangle is $1 - D_A$ \cite{Oke82} (the original
spectrum smoothed by a top hat of width $\sigma_{\rm s} = N_{\rm
p}$~pixels). Since we are only interested in local fluctuations of the
transmission around the mean, we then subtract out the mean as
calculated on the basis of our null-hypothesis. Essentially, this
removes the global redshift evolution of the optical depth. The
statistical significance of any remaining residual fluctuations around
zero are then assessed in terms of the expected rms as a function of
wavelength and scale.

In the rest of this section we calculate the relevant quantities. The
work presented here is developed from earlier calculations carried out
by \citeN{ZuoPhin93}, \citeN{Zuo93}, and \citeN{Zuo94} (but see also
\citeNP{Press93}). For completeness and clarity we reiterate some of
their derivations here. When considering the expected mean
transmission and its variance it is helpful to introduce the concept
of transmission probability. The idea is to view a \lya\ forest
spectrum as a random stochastic process \cite{Press93}. Every point in
the spectrum is a random variable, e$^{-\tau}$, drawn from the
transmission probability density function $f_\lambda({\rm
e}^{-\tau})$, also known as flux decrement distribution function
(\citeNP{Rauch97}; \citeNP{Kim97}) or distribution of intensities
(\citeNP{Jenkins91}; \citeNP{Webb92}). In principle, we have a
different probability density function at each wavelength such that
e.g.\ the moments of $f_\lambda$ are functions of wavelength. There is
a small and subtle difference between the transmission probability
density function and the distribution of pixel intensities of a
spectrum. $f_\lambda$ should in principle be measured by constructing
the frequency distribution of pixel intensities at $\lambda$ (and only
at $\lambda$) of many different spectra. Although this is important to
note we shall see later that at least the first and second moments of
$f_\lambda$ are only slowly varying functions of $\lambda$ so that in
many calculations we can approximate e$^{-\tau}$ as a stationary
stochastic process.

\subsection{The mean \lya\ transmission}
Given the distribution of absorption line parameters $\frac{d\cal
N}{dz\,dN\,db} = \eta(z,N,b)$, what is the mean transmission at a
given wavelength?  We can define an effective optical depth,
$\tau_{\rm eff}$, as a function of observed wavelength, $\lambda$, by
\be
{\rm e}^{-\tau_{\rm eff}(\lambda)} \equiv \, \langle {\rm e}^{-\tau(\lambda)} 
\rangle.
\ee
In the following we will neglect any contribution to $\tau_{\rm eff}$
from the classical Gunn-Peterson effect which is limited to $\tau_{\rm
GP} \la 0.04$ \cite{Webb92}. If the number of absorption lines per
sight-line is Poisson distributed with a mean of $m = \int_0^{\infty}
\int_0^{\infty} \int_{z_1}^{z_2} \eta(z,N,b) \,dz\,dN\,db$ then we
have
\be 
{\rm e}^{-\tau_{\rm eff}} = \sum_{k=0}^{\infty} p(k;m) \langle
{\rm e}^{-\tau_{\rm s}}\rangle^k, 
\ee 
where $p(k;m) = {\rm e}^{-m}\,m^k/ \, k!$ and 
\be
\langle {\rm e}^{-\tau_{\rm s}(\lambda)} \rangle \,=
\int_0^{\infty}\int_0^{\infty} \int_{z_1}^{z_2}
\frac{\eta(z,N,b)}{m} \: {\rm e}^{-\tau_{\rm s}(\lambda_z; \,N,b)}
\,dz\,dN\,db.  
\ee 
$\tau_{\rm s}(\lambda_z; \,N,b)$ is the profile of a single absorption
line at $z,N,b$ where $\lambda_z = \lambda / (1+z)$.  After some
algebra we find
\bea 
\tau_{\rm eff} =
m(1-\langle {\rm e}^{-\tau_{\rm s}} \rangle) &=& \int \eta (1-{\rm e}^{-\tau_{\rm
s}(\lambda_z)}) \:dz\,dN\,db \\ &=& \lambda
\int_0^{\infty}\int_0^{\infty} \int_{\lambda_{z_2}}
^{\lambda_{z_1}} \frac{\eta}{\lambda_z^2} (1-{\rm e}^{-\tau_{\rm
s}(\lambda_z)}) \:d\lambda_z\,dN\,db \nonumber 
\eea
If we exclude strongly saturated and damped systems from our analysis
then $\tau_{\rm s}(\lambda_z)$ peaks sharply at $\lambda_z =
\lambda_{\alpha} = 1215.67$~\AA\ so that
\be \label{teff} 
\tau_{\rm eff} \simeq
\frac{1+z_{\rm abs}}{\lambda_{\alpha}} \int\!\!\int \eta(z_{abs},N,b)
\int_{\lambda_{z_2}}^{\lambda_{z_1}} (1-{\rm e}^{-\tau_{\rm
s}(\lambda_z)}) \,d\lambda_z \:dN\,db, 
\ee 
where $z_{\rm abs} = \lambda / \lambda_{\alpha} - 1$. Usually, $z_1 =
\lambda_{\beta}/\lambda_{\alpha} (1+z_{\rm em})-1$, where
$\lambda_{\beta} = 1025.72$~\AA, and in the absence of a proximity
effect $z_2 = z_{\rm em}$.  For $\lambda$ close to
$\lambda_{\alpha}(1+z_{\rm em})$, there are fewer than average
absorption lines longward of $\lambda$. This produces an `edge
effect', superimposed on the well-known proximity effect
\cite{Weymann81,Cooke97}. Similarly, there will be a reverse edge
effect for $\lambda$ close to $\lambda_{\beta}(1+z_{\rm em})$ because
of the additional absorption by Ly$\beta$ lines. If $\lambda$ falls
well away from these limits then we can extend the upper and lower
integration limits in \mref{teff} to $\infty$ and $0$ respectively
because if $\lambda_z$ and $\lambda_{\alpha}$ are sufficiently far
apart $1-{\rm e}^{-\tau_{\rm s}}$ is zero. Thus we have
\be 
\tau_{\rm eff} \simeq \frac{1+z_{\rm
abs}}{\lambda_{\alpha}} \int \eta(z_{\rm abs}, N, b)\, W(N,b)
\,dN\,db.  
\ee 

Observationally $\eta$ is found to be of the form $\eta(z,N,b) =
(1+z)^{\gamma} F(N,b)$ (\citeNP{Kim97}; \citeNP{Lu96};
\citeNP{Bechtold94}; \citeNP{Williger94}; \citeNP{Bahcall93}). We
therefore arrive at
\be \label{taueff}
\tau_{\rm eff} = B(1+z_{\rm abs})^{\gamma+1} = B
\left(\frac{\lambda}{\lambda_{\alpha}}\right)^{\gamma +1}, \ee where
\be \label{B}
B = \frac{1}{\lambda_{\alpha}} \int_0^{\infty} \int_0^{\infty}
F(N,b) \, W(N,b) \,dN\,db.  
\ee
In practice, we compute $B$ directly from the data for reasons
described in section \ref{simulations}.  Thus we have \be \langle {\rm
e}^{-\tau}
\rangle \, = {\rm e}^{-B
\left(\frac{\lambda}{\lambda_{\alpha}}\right)^{\gamma +1}}. 
\ee 

\subsection{The auto-covariance} \label{auto}
The auto-covariance function of the transmission is given by
\bea
\gamma_{{\rm e}^{-\tau}}(\lambda, \lambda')
&=& \left \langle \left({\rm e}^{-\tau(\lambda)} - 
\langle {\rm e}^{-\tau(\lambda)} \rangle\right)
\left({\rm e}^{-\tau(\lambda')} - 
\langle {\rm e}^{-\tau(\lambda')} \rangle\right) 
\right \rangle \nonumber\\
&=& \langle {\rm e}^{-\tau(\lambda)} {\rm e}^{-\tau(\lambda')} \rangle - 
{\rm e}^{-\tau_{\rm eff}(\lambda)} {\rm e}^{-\tau_{\rm eff}(\lambda')} \\
&\equiv& {\rm e}^{-\Pi(\lambda, \lambda')}-{\rm e}^{-\tau_{\rm eff}(\lambda)} 
{\rm e}^{-\tau_{\rm eff}(\lambda')}. \nonumber
\eea
Following the same calculations as in the previous section, we find
\be \label{Pi}
\Pi(\lambda, \lambda') = \int \eta(z, N, b) \,
(1-{\rm e}^{-\tau_{\rm s}(\lambda)} {\rm e}^{-\tau_{\rm s}(\lambda')})
\,dz\,dN\,db.
\ee
Let us consider the variance of the transmission given by
\be
\sigma_{{\rm e}^{-\tau}}^2 = \gamma_{{\rm e}^{-\tau}}(\lambda, \lambda).
\ee
Since $\tau_{\rm s}(N) \propto N$, we have $2\tau_{\rm s}(N) =
\tau_{\rm s}(2N)$ and thus we get similarly to equation \mref{taueff}
\be
\Pi(\lambda, \lambda) = \tilde B (1+z)^{\gamma +1},
\ee
where
\be
\tilde B = \frac{1}{\lambda_{\alpha}}
\int_0^{\infty} \int_0^{\infty} F(N,b) \, W(2N,b) \,dN\,db.
\ee
Observations have shown that the distribution of column densities can
be fairly well represented by a power law, $F(N,b) = N^{-\beta} f(b)$
\cite{Carswell84} with $\beta \approx 1.5$ (\citeNP{Kim97};
\citeNP{Kirkman97}; \citeNP{Lu96}; \citeNP{Hu95}). A finite number of
absorption lines per line of sight implies that the power law must
break off at the low $N$ end at some $N_{\rm low}$.  We also expect a
break at the high $N$ end at some $N_{\rm hi}$.  Thus we have
\be
\tilde B = \frac{2^{\beta-1}}{\lambda_{\alpha}}
\int_0^{\infty} \int_{2 N_{\rm low}}^{2 N_{\rm hi}} F(N,b) 
\, W(N,b) \,dN\,db.
\ee
We know that the power law is a good approximation for the range $12
\la \log N/{\rm cm}^{-2} \la 22$ (\citeNP{Hu95};
\citeNP{Petitjean93}), so that $N_{\rm low}$ and $N_{\rm hi}$ are in
the linear and square-root regimes of the curve of growth
respectively. Under this assumption it is straightforward to show that
$\tilde B$ can be well approximated by $2^{\beta-1}B$ for $\beta \la
1.8$, this being the exact result (for all $\beta$) if there are no
breaks in the power law. Therefore we finally arrive at
\be \label{sigmaetau}
\sigma_{{\rm e}^{-\tau}}^2 = {\rm e}^{-2^{\beta-1} B (1+z)^{\gamma+1}} - 
{\rm e}^{-2 B (1+z)^{\gamma+1}}.
\ee

\subsection{Instrumental effects} \label{instrument}
So far we have not considered any instrumental effects. There are two
classes of such effects: finite spectral resolution and various
sources of noise.
\subsubsection{Finite resolution} \label{resol}
A new stochastic variable $X$ is produced by convolving e$^{-\tau}$
with a line spread function (LSF) $L$:
\be
X(\lambda) = \int_{-\infty}^{\infty} {\rm e}^{-\tau(\lambda')} 
L(\lambda - \lambda') \, d\lambda'
\ee
For the mean of $X$ we get
\be
\langle X \rangle \, = \int_{-\infty}^{\infty} 
\langle {\rm e}^{-\tau(\lambda')} \rangle 
L(\lambda - \lambda') \, d\lambda'.
\ee
Although it has been stressed that the mean and the variance of
e$^{-\tau}$ are functions of $\lambda$ we will now approximate
e$^{-\tau}$ as a {\em stationary} stochastic process because both the
mean and the variance are smooth, slowly varying functions of
$\lambda$.  \citeN{Lu94} have shown this approximation to be valid.
Thus we have
\be \label{meanX}
\langle X \rangle(\lambda) \simeq {\rm e}^{-\tau_{\rm eff}(\lambda)}
\ee
since the LSF is normalised to unity. As is intuitively clear, the
convolution does not change the mean transmission.

The auto-covariance function of $X$ is given by
\bea
\gamma_X(\lambda_1, \lambda_2)
&=& \langle \left(X(\lambda_1) - \!\langle X \rangle(\lambda_1)\right) \:
\left(X(\lambda_2) - \!\langle X \rangle(\lambda_2)\right) \rangle \nonumber\\
&=& \int \limits_{-\infty}^{+\infty} \!\!\! \int L(\lambda_1 - \lambda_1') \,
L(\lambda_2 - \lambda_2') \, \gamma_{{\rm e}^{-\tau}}(\lambda_1',\lambda_2')
\, d\lambda_1' \, d\lambda_2'.
\eea
Since we consider e$^{-\tau}$ to be a {\em stationary} process,
$\gamma_{{\rm e}^{-\tau}}$ depends only on $u' = \lambda_2' -
\lambda_1'$ and $\gamma_X$ depends only on $u = \lambda_2 - \lambda_1$
\cite{Jenkins}. Usually, the LSF can be well approximated as a
Gaussian. After some algebra we get
\be \label{gX}
\gamma_X(u) = \frac{1}{\sqrt{2\pi}\sigma_{\rm LSF}'}
\int_{-\infty}^{\infty} \gamma_{{\rm e}^{-\tau}}(u') \,
\exp \left(-\frac{(u - u')^2}{2\sigma_{\rm LSF}'^2} \right) \,du',
\ee
where $\sigma_{\rm LSF}' = \sqrt{2} \, \sigma_{\rm LSF}$.
  
\subsubsection{Noise}
The noise in optical spectra is mainly due to photon counting
statistics, detector read-out noise, dark current, sky subtraction,
and cosmic rays. As the Poisson statistics of the absorption lines are
expected to dominate the variance we have not attempted to model the
noise characteristics in great detail. We rather approximate the
cumulative effect of all the noise components mentioned above to be
Gaussian. Therefore, we define the stochastic variable $Y$ by
\be \label{addnoise}
Y(\lambda) = X(\lambda) + n(X(\lambda)),
\ee
where $n$ is a random variable drawn from a Gaussian with mean zero
and variance $\sigma_n^2(X) = X\,(c_1 - c_2) + c_2$. The constants
$c_1$ and $c_2$ characterise the photon counting statistics and the
sky subtraction plus detector noise ($c_1 > c_2$). For the mean of $Y$
we have
\be
\langle Y \rangle \,=\, \langle X \rangle + \langle n \rangle 
\,=\, \langle X \rangle \,=\, \langle {\rm e}^{-\tau} \rangle
\ee
and the covariance is given by
\be
\gamma_Y(\lambda_1, \lambda_2) =
\gamma_X(\lambda_1, \lambda_2) + \gamma_{Xn}(\lambda_1, \lambda_2)
+\gamma_{Xn}(\lambda_2, \lambda_1) + \gamma_n(\lambda_1, \lambda_2).
\ee
$\gamma_{Xn}$ denotes the {\em cross}-covariance function of $X$ and
$n$.  Although $X$ and $n$ are not {\em independent} they are, by
construction, {\em uncorrelated}, so that $\gamma_{Xn} = 0$.
\citeN{Zuo94} showed that the originally uncorrelated photon noise in
different wavelength bins remains uncorrelated after passing through a
spectrograph of finite resolution. Therefore $\gamma_n(u)$ must be
discontinuous at $u = 0$:
\be
\gamma_n(u) = \left \{
\begin{array}{ll}
0 & u > 0 \\
\int f_X(x) \, \sigma_n^2(x) \, dx & u = 0
\end{array} \right.
\ee
where $f_X(x)$ denotes the pdf of $X$. The integral reduces to 
$\sigma_n^2(\langle X \rangle)$. Thus
\be \label{gY}
\begin{array}{rcll}
\sigma_Y^2 &=& \sigma_X^2 + \sigma_n^2(\langle X \rangle) &\\
\gamma_Y(u) &=& \gamma_X(u) & u > 0.
\end{array}
\ee

\subsection{Filter matching} \label{G}
In order to develop a method for detecting structures of arbitrary
scale, we proceed by convolving the spectrum with a smoothing function
of smoothing scale $\sigma_{\rm s}$.  The convolution filters out all
power on scales smaller than $\sigma_{\rm s}$. By changing the width
of the smoothing function we can match the filter width to the scale
of any feature and thus maximise its signal. In practice, we perform
the convolution successively at all possible smoothing scales. At the
largest possible scale ($\sigma_{\rm s, max} =$ number of pixels in
the spectrum) the entire spectrum is compressed into a single number
whereas on the smallest scale ($\sigma_{\rm s, min} = 1$~pixel) the
spectrum remains essentially unchanged. These two extremes correspond
to the tip and the base of the triangle which forms when the
successive convolutions of the spectrum are plotted in the ($\lambda,
\sigma_{\rm s}$) plane. In principle, there are many choices for the
specific form of the smoothing function but for simplicity we will use
a Gaussian, thus constructing a new stochastic variable $G$:

\be
G(\lambda, \sigma_{\rm s}) = \frac{1}{\sqrt{2\pi} \; \sigma_{\rm s}}
\int_{-\infty}^{\infty} Y(\lambda') \exp(-\frac{(\lambda-\lambda')^2}
{2\sigma_{\rm s}^2}) d\lambda'.
\ee
As in section \ref{resol} we have
\be \label{mG}
\langle G \rangle (\lambda) \simeq {\rm e}^{-\tau_{\rm eff}(\lambda)}
= {\rm e}^{-B (\frac{\lambda}{\lambda_\alpha})^{\gamma+1}}.
\ee
Note that the use of a top hat function would yield a variable akin to
$1-D_A$ (and the same result as equation \mref{mG}), where $D_A$ is
the flux deficit parameter \cite{Oke82}. The observations
are consistent with this result (\citeNP{Press93}; \citeNP{ZuoLu93};
but see also \citeNP{Bi97}). Similar to equation \mref{gX} we find
\bea \label{gG}
\gamma_G(u) &=& \frac{1}{\sqrt{2\pi} \; \sigma_{\rm s}'} 
\int_{-\infty}^{\infty}
\gamma_Y(u') \exp(-\frac{(u-u')^2}{2\sigma_{\rm s}'^2}) du' \nonumber\\
&=& \frac{\sigma_n^2({\rm e}^{-\tau_{\rm eff}})}{\sqrt{2\pi} \; 
\sigma_{\rm s}'/ps}
\, {\rm e}^{-\frac{u^2}{2\sigma_{\rm s}'^2}} + \frac{1}{\sqrt{2\pi} \,
\sqrt{\sigma_{\rm s}'^2 + \sigma_{\rm LSF}'^2}} \int_{-\infty}^{\infty}
\gamma_{{\rm e}^{-\tau}}(u'') \exp(-\frac{(u-u'')^2}
{2(\sigma_{\rm s}'^2 + \sigma_{\rm LSF}'^2)}) \; du''
\eea
where $ps$ denotes the pixel size in \AA.  To proceed we need to
consider the auto-covariance function of a `perfect' spectrum,
$\gamma_{{\rm e}^{-\tau}}$, in more detail. In principle, it can be
calculated from equation \mref{Pi} as was done by \citeN{Zuo94} for a
single Doppler parameter rather than a distribution of $b$ values. The
result is a rather unwieldy numerical integral. Here we can take a
different approach.  As expected, we can see from equation \mref{gG}
that the quantity that we are interested in, $\sigma_G^2 =
\gamma_G(0)$, does {\em not} depend on the exact shape of
$\gamma_{{\rm e}^{-\tau}}$ but rather on the convolution of
$\gamma_{{\rm e}^{-\tau}}$ with a Gaussian. We may therefore hope to
be able to use a simpler analytic approximation for $\gamma_{{\rm
e}^{-\tau}}$ since all systematic differences will be somewhat `washed
out' by the convolution. Ultimately, this procedure must be justified
by its success. We shall return to this point when we compare the
results of this section to simulations. The most obvious (because
simplest) approximation for $\gamma_{{\rm e}^{-\tau}}$ is a Gaussian,
especially when considering that unsaturated Voigt profiles are very
nearly Gaussian:
\be \label{gYapp}
\gamma_{{\rm e}^{-\tau}}(u) \simeq \sigma_{{\rm e}^{-\tau}}^2 \; 
{\rm e}^{-\frac{u^2}{2q^2}}.
\ee
Since we are operating in wavelength space rather than in velocity
space the width, $q$, must be a function of wavelength, because an
absorption line with a given Doppler parameter will be wider in wavelength
space at higher redshifts than at lower redshifts. This is of course
just another reflection of the fact that ${\rm e}^{-\tau}$ is not a
stationary process. But again, $q$ will vary only slowly with
wavelength (approximately linearly) so that the stationary
approximation is valid. Using this approximation we find
\be \label{sigG}
\sigma_G^2(\lambda, \sigma_{\rm s}) = 
\frac{\sigma_n^2({\rm e}^{-\tau_{\rm eff}(\lambda)})}
{2\sqrt{\pi} \; \sigma_{\rm s}/ps} + \frac{\sigma_{{\rm e}^{-\tau}}^2(\lambda)}
{\sqrt{2\frac{\sigma_{\rm s}^2 + \sigma_{\rm LSF}^2}{q^2(\lambda)} + 1}}.
\ee  
Equations \mref{mG} and \mref{sigG} are the final result of this section.

\section{Simulations} \label{simulations}
The motivation for simulations of \lya\ forest spectra in this work is
threefold. First of all we need to determine the parameters $B$ and
$q$.  The normalisation $B$ could be calculated numerically from
equation \mref{B}. However, it is clear that for real data small
inaccuracies in the zeroth and first order of the continuum fit will
cause an artificial offset of the measured mean transmission from the
calculated one. In anticipation of this problem we choose to determine
$B$ directly from the data.  Since equation \mref{gYapp} is only an
approximation we cannot {\em a priori} calculate a precise value for
$q$. We therefore have to measure it from simulations.  Secondly, we
would like to check the validity of equations \mref{mG} and
\mref{sigG} by comparing the calculations with an analysis of
simulated spectra. Thirdly, we would like to compare the sensitivity
of the new analysis to the presence of non-random structures to that
of the traditional line counting technique. In order to cater for this
third need, we employed a more sophisticated method than simply
randomly drawing the parameters of absorption lines from a given
distribution $\eta(z,N,b)$. Instead we distribute absorbers in a
cosmological volume and take lines of sight through that volume. This
provides the flexibility of introducing specific types of clustering
models. We assume absorbers to be spherical and prescribe a column
density -- impact parameter relationship of the form $N(r) = N_0
(r/r_0)^{-a}$ which has been observed at low redshift where galaxies
are unambiguously associated with \lya\ absorbers (\citeNP{Chen98};
\citeNP{Lanzetta96}; \citeNP{Lanzetta95}; but see also
\citeNP{Bowen96}). This procedure simply ensures that the column
density distribution of the absorption lines will be of the form
$N^{-\beta}$ with $\beta = 2/a + 1$. We draw Doppler parameters from a
truncated Gaussian. We choose to keep the comoving number density of
absorbers constant and thus ascribe their redshift evolution solely to
the evolution of their absorption cross-section. This requires a
redshift dependence of $r_0$
\be
 r_0(z) = r_0(\hat z)\:\frac{(1+2 q_0 z)^\frac{1}{4} \:
(1+z)^\frac{\gamma-1}{2}}{(1+q_{0})^\frac{1}{4} \: (1 + \hat z)
^\frac{\gamma-1}{2}} 
\ee 
where we take the normalisation $r_0(\hat z) = 1~h^{-1}$~Mpc at
$N_0 = 10^{12}$~cm$^{-2}$ from \citeN{Lanzetta95} at $\hat z =
0.5$. 

\subsection{$B$ and $q$ from simulations}
\begin{figure}
\psfig{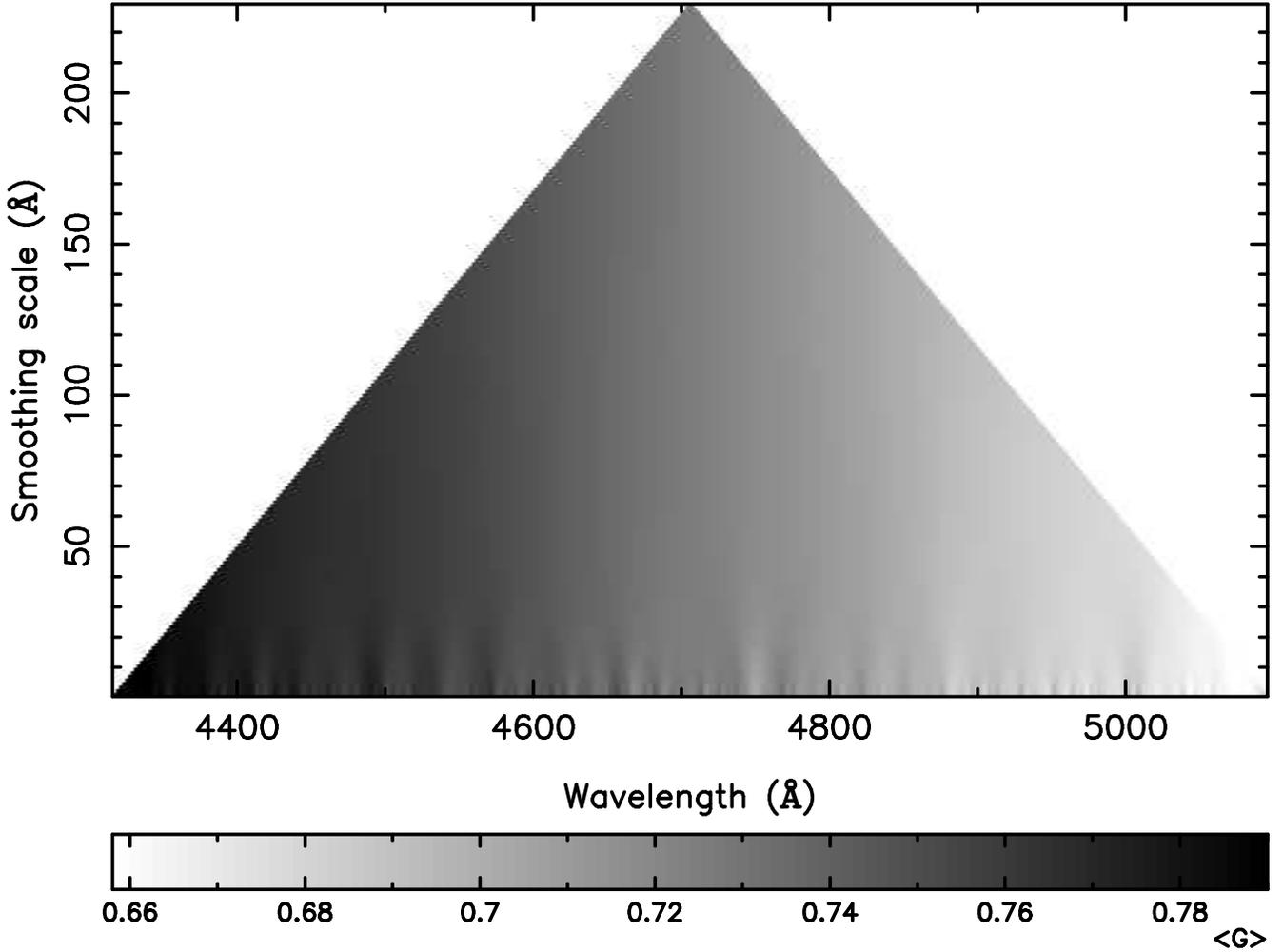}
\caption{Mean transmission triangle produced from 1000 simulated spectra.}
\label{mtrans}
\end{figure}

\begin{figure}
\psfig{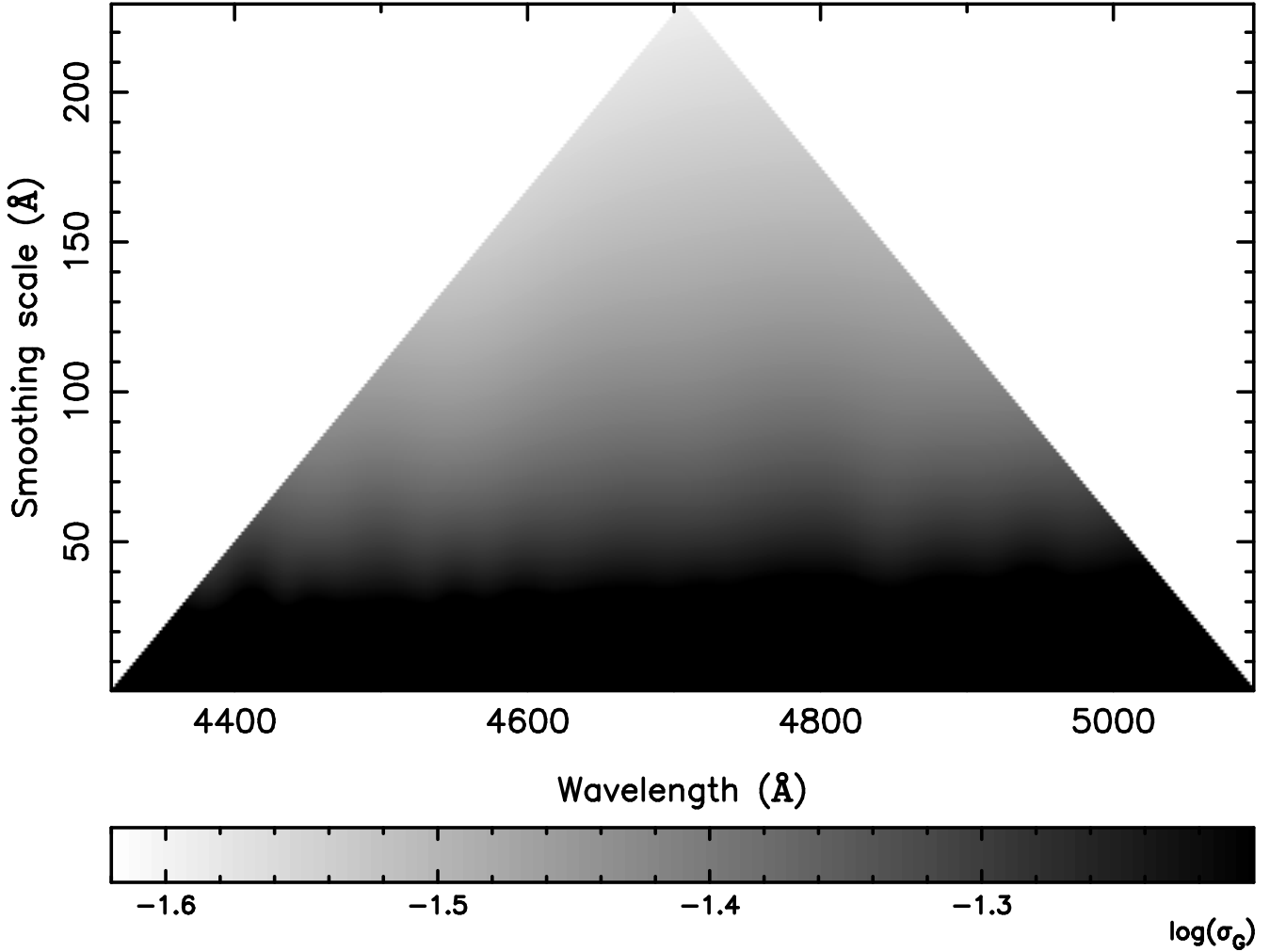}
\caption{Rms transmission triangle produced from 1000 simulated spectra.}
\label{rmstrans}
\end{figure}

\begin{figure}
\begin{center}
\leavevmode
\psfig{file=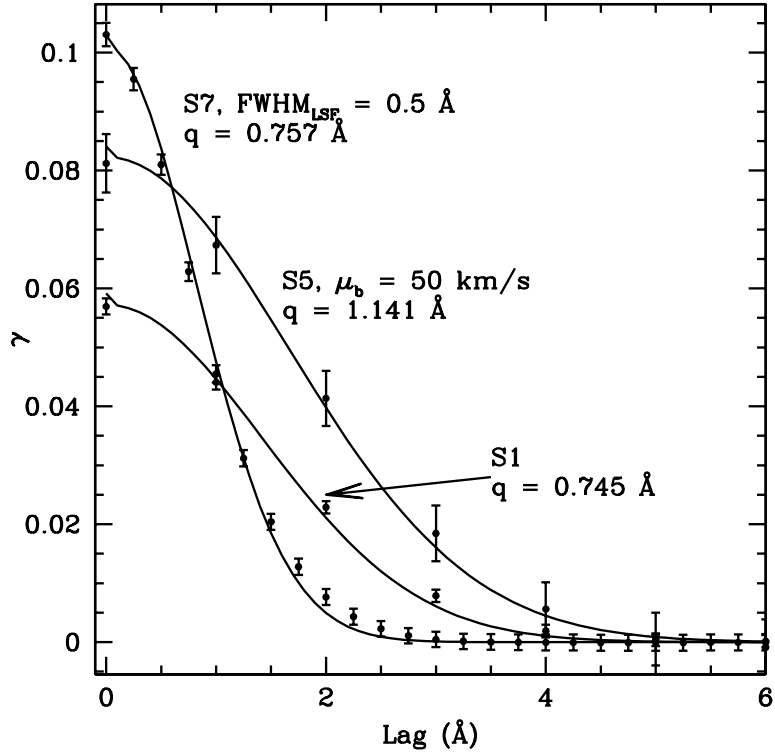,width=0.6\textwidth,silent=}
\end{center}
\caption{Auto-covariance functions of different models as indicated
and best fits (solid lines). For clarity, all errorbars have been
exaggerated by a factor of 10.}
\label{cov}
\end{figure}

In order to compare equations \mref{mG} and \mref{sigG} with
simulations we have produced a set of 1000 spectra in the manner
described in the previous section with randomly distributed
absorbers. The spectra are convolved with a line spread function and
noise is added according to equation \mref{addnoise}. The parameters
of the simulation are listed in Table \ref{paratab} (S1). For each
spectrum we constructed its transmission triangle using a Gaussian
smoothing function. From this set of 1000 triangles we produced the
mean and rms transmission triangles which are shown in Figures
\mref{mtrans} and \mref{rmstrans}.
\begin{table*}
\begin{minipage}{110mm}
\caption{Parameters of simulations.}
\label{paratab}
\begin{tabular}{@{}lcccccccc}
\hline
& $\gamma$ & $\beta$ & $n_0$ & $\mu_b$ & $\sigma_b$
& $b_{\rm cut}$ & S/N & FWHM$_{\rm LSF}$\\
& & & ($h^3$ Mpc$^{-3}$) & (km s$^{-1}$) & (km s$^{-1}$) & (km s$^{-1}$)
& & (\AA)\\
\hline
S1 & 2.5 & 1.5 & 0.01 & 30 & 8 & 18 & 20 & 2\\
S2 & 2.5 & 1.7 & 0.01 & 30 & 8 & 18 & 20 & 2\\
S3 & 2.7 & 1.5 & 0.01 & 30 & 8 & 18 & 20 & 2\\
S4 & 2.5 & 1.5 & 0.015 & 30 & 8 & 18 & 20 & 2\\
S5 & 2.5 & 1.5 & 0.01 & 50 & 8 & 38 & 20 & 2\\
S6 & 2.5 & 1.5 & 0.01 & 30 & 16 & 18 & 20 & 2\\
S7 & 2.5 & 1.5 & 0.01 & 30 & 8 & 18 & 20 & 0.5\\
S8 & 2.5 & 1.5 & 0.01 & 30 & 8 & 18 &  5 & 2\\
\hline
\end{tabular}
$n_0$ is the comoving number density of absorbers (normalisation of
$\eta(z,N,b)$), $\mu_b$, $\sigma_b$, and $b_{\rm cut}$ are the mode,
width, and lower cut-off of the Doppler parameter distribution
respectively. For models S1 and S7 we created 1000 spectra, in all
other cases we simulated 100 spectra. For all spectra $\langle z
\rangle = 2.87$.
\end{minipage}
\end{table*}
Before we can go on to compare these results with equations \mref{mG}
and \mref{sigG} we must determine the values of the two parameters $B$
and $q$. We fix the normalisation $B$ at the tip of the mean
transmission triangle by requiring
\be
\langle G \rangle (\sigma_{\rm s, max}, \lambda_{\rm c}) =
{\rm e}^{-B(\frac{\lambda_{\rm c}}{\lambda_\alpha})^{\gamma+1}},
\ee
where $\sigma_{\rm s, max}$ denotes the biggest possible smoothing scale
and $\lambda_{\rm c}$ is the central wavelength of the region under
consideration. Having stipulated equation \mref{gYapp} we measure $q$
(at $\lambda_{\rm c}$) from the simulations by performing a single
parameter $\chi^2$ fit of the function
\be \label{gYapp2}
\gamma_Y(u) = \left \{
\begin{array}{ll}
\sigma_n^2({\rm e}^{-\tau_{\rm eff}(\lambda_{\rm c})}) +
\frac{\sigma_{{\rm e}^{-\tau}}^2(\lambda_{\rm c})}{\sqrt{2 
\frac{\sigma_{\rm LSF}^2}{q^2} + 1}} & u = 0 \\
\frac{\sigma_{{\rm e}^{-\tau}}^2(\lambda_{\rm c})}{\sqrt{2 
\frac{\sigma_{\rm LSF}^2}{q^2} + 1}} \exp(\frac{-u^2}
{2 (2 \sigma_{\rm LSF}^2 + q^2)}) & u > 0
\end{array} \right.
\ee
to the mean auto-covariance function of the 1000 simulated
spectra. Since equation \mref{gYapp} (and hence equation
\mref{gYapp2}) is an approximation we do not {\it a priori} expect a
statistically acceptable fit. Nevertheless, in practice this procedure
provides a reliable estimate of $q$ because both the shape (width)
{\em and} normalisation of $\gamma_Y$ are sensitive to $q$. Figure
\mref{cov} shows the measured mean auto-covariance function of S1 and
its fit. The same is also plotted for two other sets of spectra
(c.f.~Table \ref{paratab}), S7 (same model as S1 but the spectra are
of higher resolution) and S5 (larger mode of the Doppler parameter
distribution).
It is evident that a Gaussian does not adequately represent the
auto-covariance functions; a Gaussian has too much power
on small scales and too little power on larger scales. Indeed, each
fit produces an unacceptably large $\chi^2$, although we point out
that in any case a somewhat larger than usual $\chi^2$ must be
anticipated because of the non-Gaussian and correlated nature of the
measurement errors of $\gamma_Y$. However, we recall that we are
mostly interested in the typical width and strength of the correlation
rather than its exact shape. Since both sets of spectra S1 and S7
should yield the same value for $q$, the purpose of set S7 was to
check whether the above method of determining $q$ is robust and to
provide an estimate of the true error in $q$ as opposed to the formal
error as calculated from the $\chi^2$ fit. As expected, $q$ is of the
order of the mode Doppler parameter, $\mu_b$, for a range of sensible
values for $\mu_b$, as seen from S5. In fact, $q$ is seen to vary
almost linearly with $\mu_b$, which justifies $q(\lambda) =
q(\lambda_{\rm c}) \lambda / \lambda_{\rm c}$. We have also
investigated the behaviour of $q$ as a function of the other model
parameters and have found, as expected, that $q$ is only sensitive to
the parameters of the Doppler parameter distribution, $\mu_b$ and
$\sigma_b$, and of the column density distribution, $\beta$. It is
insensitive to the redshift evolution, overall normalisation, and the
quality of spectra since the $q$ values measured from models S3, S4,
S7, and S8 are all comparable. We conclude that the error in
estimating $q$ is dominated by the errors in $\mu_b$, $\sigma_b$ and
$\beta$.

\subsection{Comparison of analytical to numerical results}
\begin{figure}
\begin{center}
\leavevmode
\psfig{file=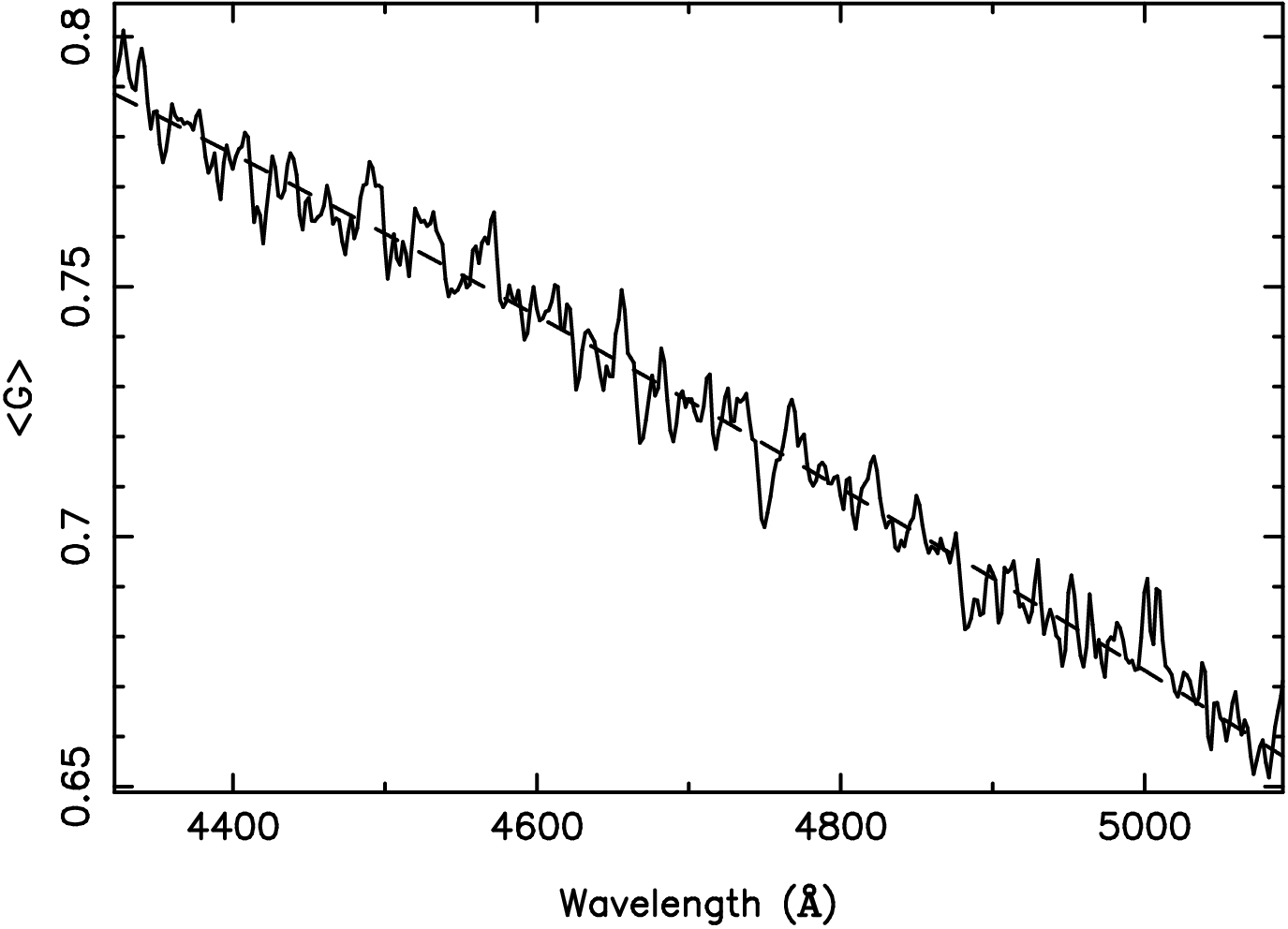,width=0.7\textwidth,angle=-90,silent=}
\end{center}
\caption{Cross section of the mean transmission triangle (c.f.~Figure
\ref{mtrans}) at FWHM$_{\rm s} = 3.2$~\AA. The dashed line shows the
prediction of equation \mref{mG}.}
\label{mGcrossw}
\end{figure}

With the values of $B$ and $q$ thus determined we can now directly
compare the results from the simulations with equations \mref{mG} and
\mref{sigG}. Figures \mref{mGcrossw} and (\ref{rmsGcross}a) show cross
sections of the mean and rms transmission triangles of S1 as functions
of wavelength at smoothing scale FWHM$_{\rm s} = 3.2$~\AA. Figure
(\ref{rmsGcross}b) shows a cross section through the rms transmission
triangle as a function of smoothing scale at $z = 2.87$. The dashed
lines show the calculations. Using the covariance matrix implied by
equation \mref{gYapp2}, a $\chi^2$ test performed on the base of the
mean transmission triangle yields $P(>\chi^2) = 0.12$ and thus the
model agrees very well with the simulations. For the rms the agreement
is not quite as good. We find that for very large smoothing scales
(FWHM$_{\rm s} > 100$~\AA) the model underestimates the rms by $\sim
4$ per cent. For smaller (and more relevant) scales the model fares
progressively better.

We have repeated this exercise for all sets of simulations listed in
Table \ref{paratab} and have always found the same good agreement. In
addition, we have repeated the calculations in section \mref{G} and
the analysis of the simulated data for the case of a top hat smoothing
function and these also agree very well. Thus we conclude that the
errors in determining any fluctuations of the \lya\ absorption around
its expected mean and in estimating their significance will be
dominated by the uncertainties in the assumed values of the parameters
$\beta$, $\mu_b$, $\sigma_b$ and to lesser extent $\gamma$ and the
overall normalisation. Any errors made in any of the approximations of
the previous sections are small compared to these uncertainties.
\begin{figure}
\begin{center}
\leavevmode
\psfig{file=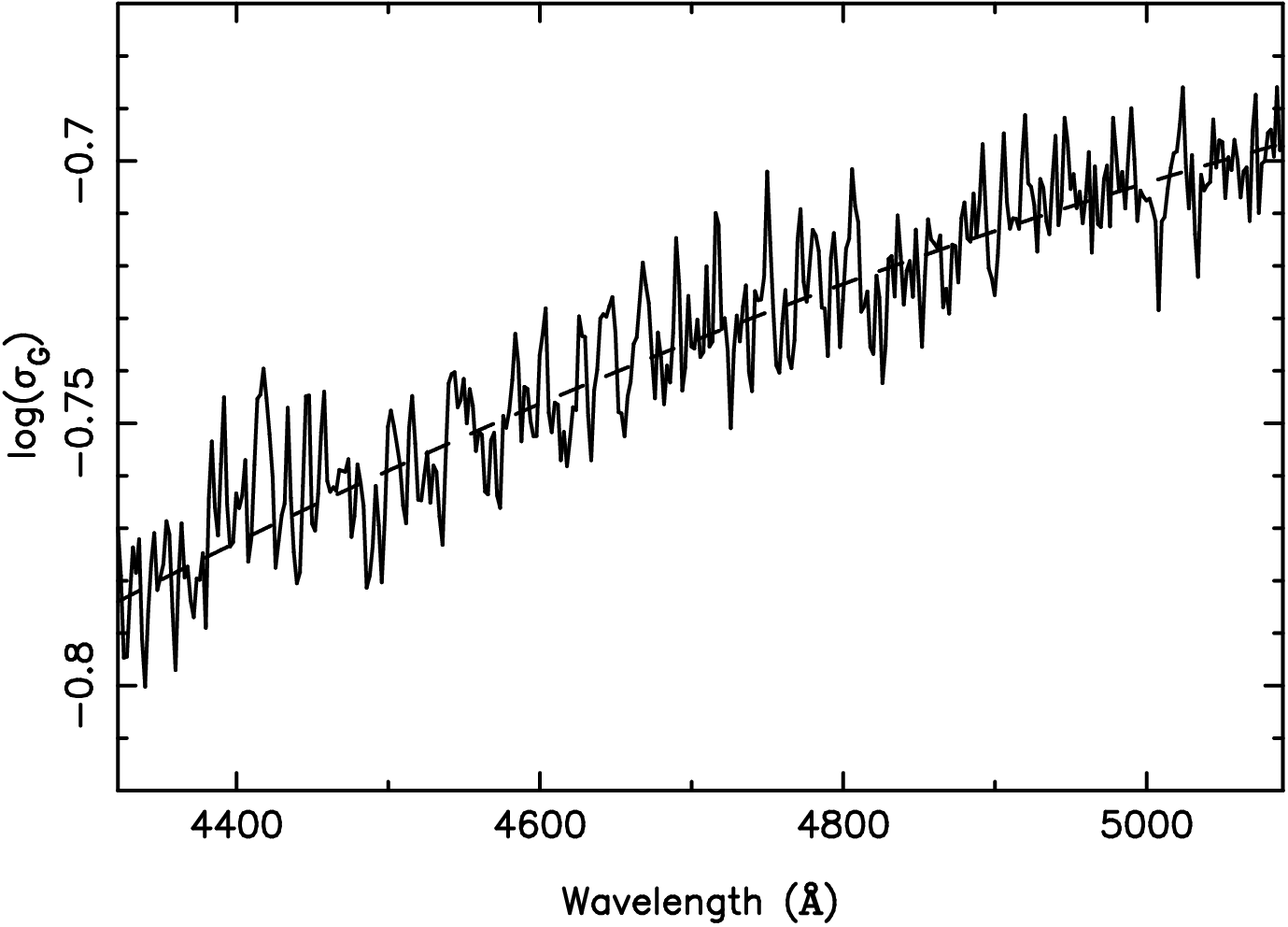,width=0.7\textwidth,angle=-90,silent=}
\end{center}
\caption{(a) Cross section of the rms transmission triangle
(c.f.~Figure \ref{rmstrans}) at FWHM$_{\rm s} = 3.2$~\AA. The dashed
line shows the prediction of equation \mref{sigG}.}
\label{rmsGcross}
\end{figure}
\begin{figure}
\begin{center}
\leavevmode
\psfig{file=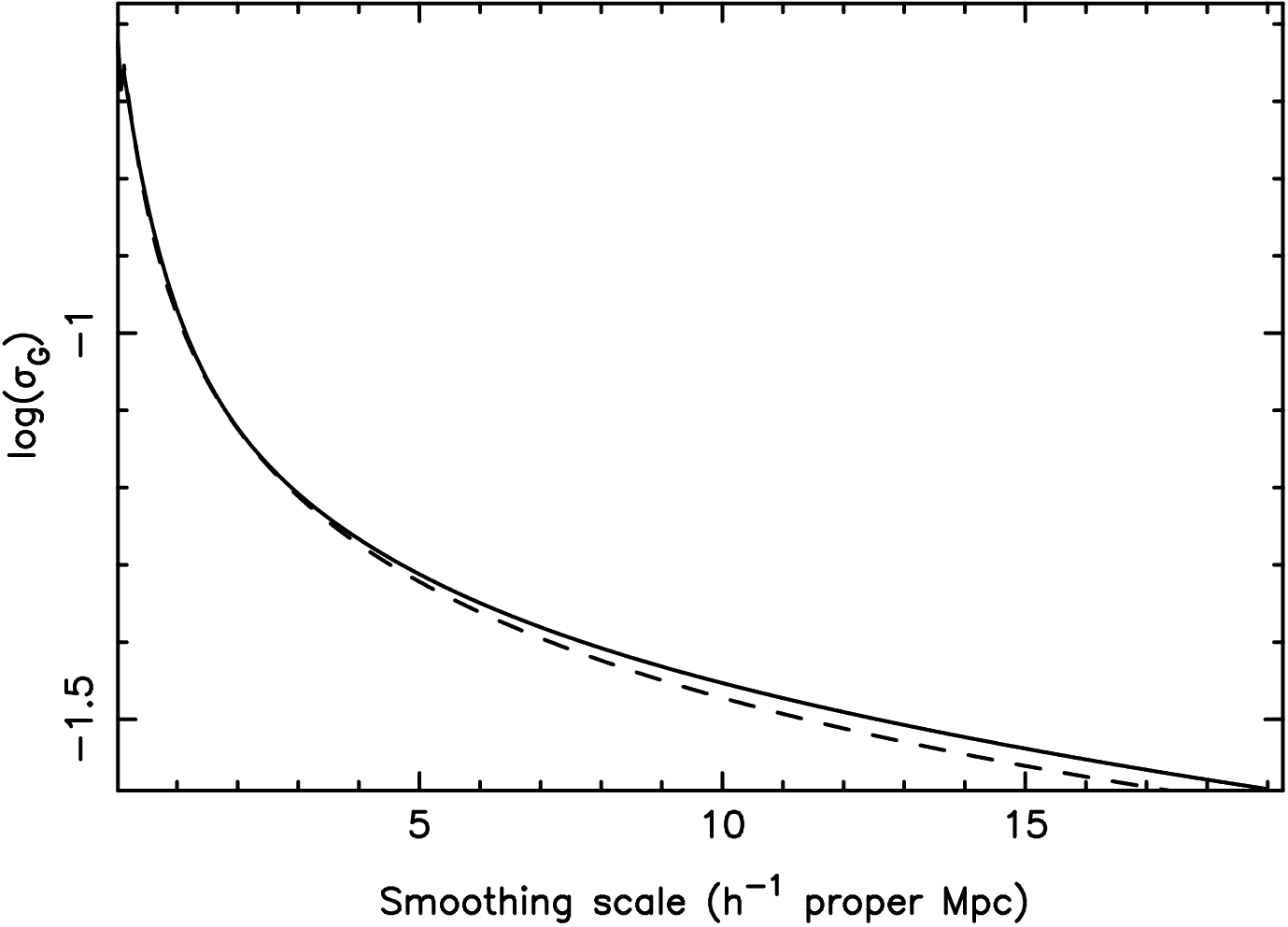,width=0.7\textwidth,angle=-90,silent=}
\end{center}
\contcaption{(b) Cross section of the rms transmission triangle
(c.f.~Figure \ref{rmstrans}) at $z= 2.87$. The dashed line shows the
prediction of equation \mref{sigG}.}
\end{figure}

\subsection{Sensitivity}
With all the calculations and parameter values in place we can now
answer the questions: `How statistically significant is an enhancement
of the local absorption line number density over the mean line number
density at redshift $z$ by a factor of $\delta n$ on the scale of
$x~h^{-1}$~Mpc?' and `At what redshift is an overdensity of $\delta n$
on the scale of $x~h^{-1}$~Mpc most significant?' To address these
questions we plot the quantity
\[
\frac{{\rm e}^{-\delta n \tau_{\rm eff}} - {\rm e}^{-\tau_{\rm
eff}}}{\sigma_G}
\]
in Figures (\ref{sensi}a) and (b) as a function of $\delta n$ and $z$
respectively for a scale of $5 \; h^{-1}$ proper Mpc (FWHM of
smoothing Gaussian), assuming the parameters of S1. 
\begin{figure}
\begin{center}
\leavevmode
\psfig{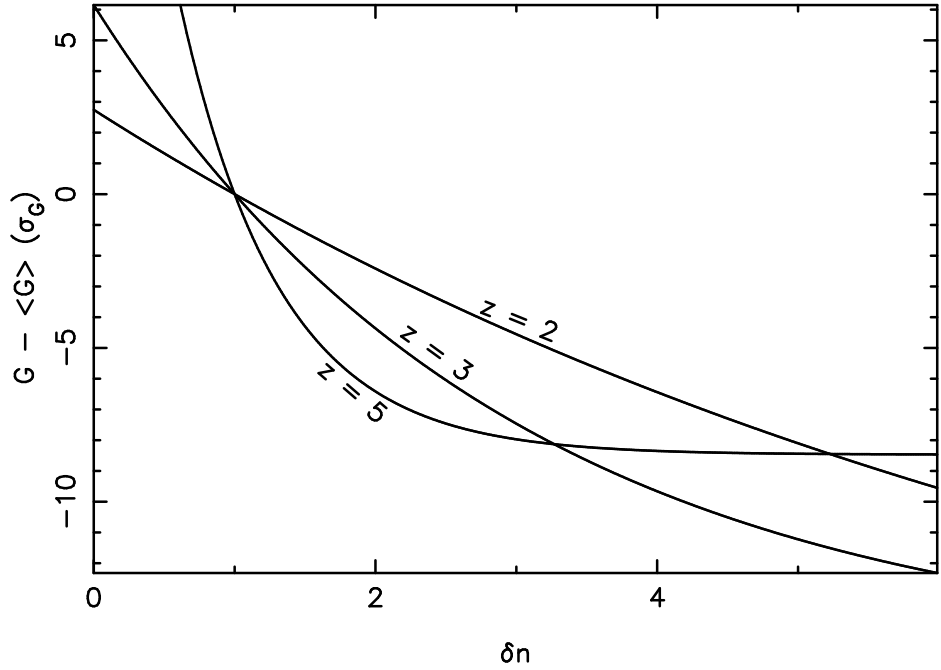}
\end{center}
\caption{(a) Expected signal in units of $\sigma_G$ of an overdensity
of absorption lines of scale $5 \; h^{-1}$ proper Mpc (FWHM of
smoothing Gaussian) at redshifts 2, 3, and 5.}
\label{sensi}
\end{figure}
\begin{figure}
\begin{center}
\leavevmode
\psfig{file=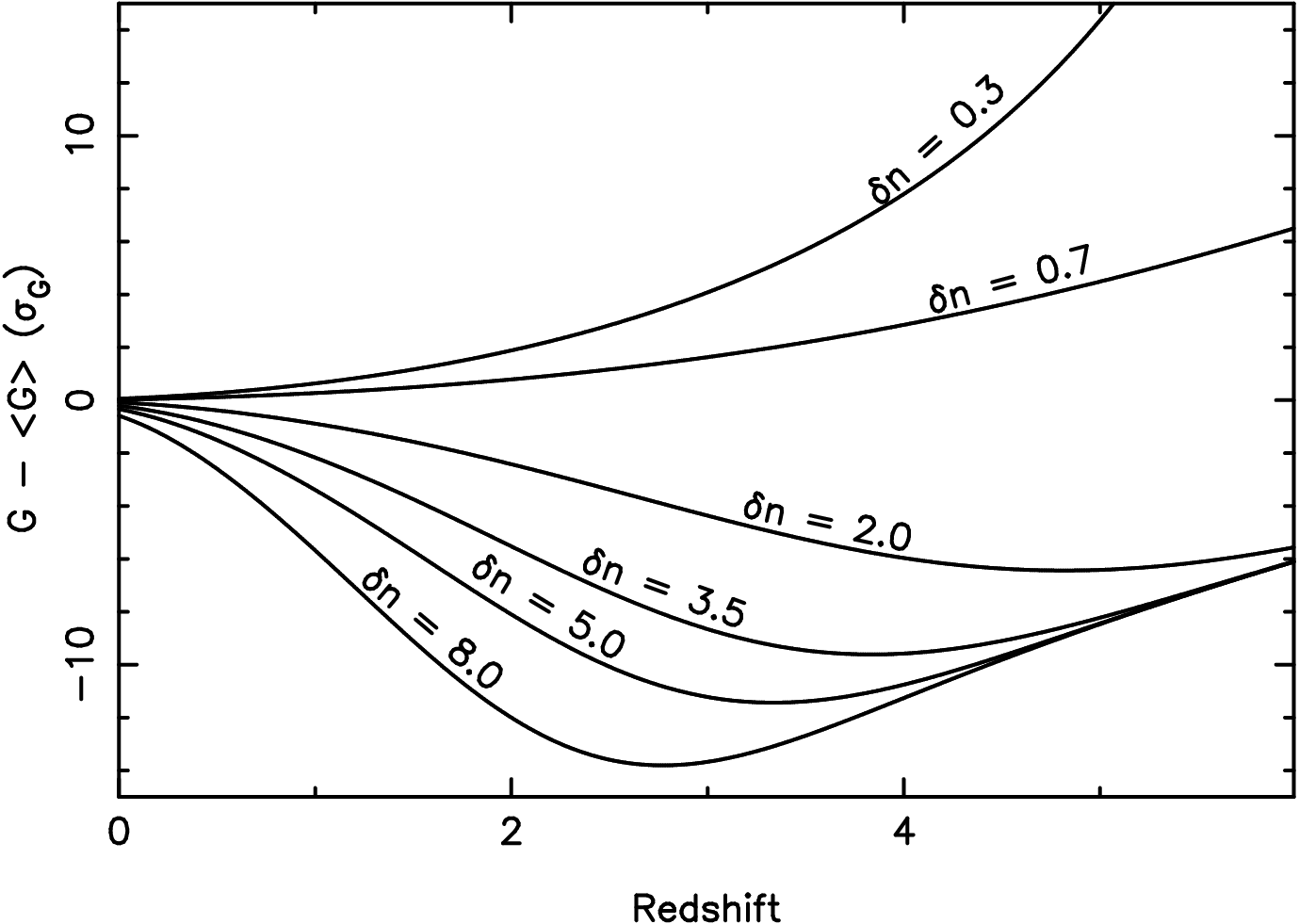,width=0.7\textwidth,angle=-90,silent=}
\end{center}
\contcaption{(b) Same as (a) as a function of redshift for the
indicated overdensities.}
\end{figure}
From Figure (\ref{sensi}a) we see that for a given redshift we can
expect a maximum signal which cannot be exceeded. This is due to
saturation as the number density of absorption lines increases rapidly
towards higher redshift. Figure (\ref{sensi}b) tells us that for a
given level of overdensity there is an optimum redshift at which this
level of overdensity will produce a maximum signal.

At this point it is necessary to comment on the exact significance of,
e.g., a `$3\sigma$ event'. For small smoothing scales the pdf of $G$
is inherently non-Gaussian such that we expect the probability of $G$
lying within $3\sigma$ of the mean to be smaller than 0.9973. In fact
the pdf is skewed such that the probability of a $+3\sigma$ event (a
void) is lower than the probability of a $-3\sigma$ event (a
cluster). At larger smoothing scales the Central Limit Theorem
guarantees Gaussianity. Thus a $3\sigma$ event at large smoothing
scales is statistically more significant than a similar event at small
smoothing scales. This additional complication must be kept in mind.

\subsection{Comparison to TPCF}
Groups of QSOs that are closely spaced in the plane of the sky can be
used to map out the large-scale 3-dimensional structure of the
intervening absorbing gas by identifying absorption features that are
approximately coincident in redshift space in two or more spectra.
One of the advantages of the analysis presented here is that it can
easily be applied to the spectra of such groups: the transmission
triangles of the different spectra are simply averaged where they
overlap. For sight-line separations of several arcminutes, different
lines of sight will not intersect the same absorber, so that according
to our null-hypothesis of an unclustered \lya\ forest different lines
of sight are uncorrelated. Therefore the variance of a mean
transmission triangle (averaged over multiple lines of sight) at
($\lambda, \sigma_{\rm s}$) is simply given by $\sigma_G^2(\lambda,
\sigma_{\rm s})$ divided by the number of triangles overlapping at
($\lambda, \sigma_{\rm s}$). Thus the signal of any structure
extending across several lines of sight will be enhanced.

In order to compare our analysis directly to a `traditional' two-point
correlation function analysis we have simulated spectra of a close
group of QSOs where the absorbers are clustered. In view of the
modern, large hydrodynamic simulations of structure formation which
reproduce many of the observed properties of the \lya\ forest, the
simulations presented here must be understood in the sense of a toy
model. The advantage of our simulation is the flexibility to model
different clustering characteristics, thus enabling us to test our
method comprehensively. It is not important for these particular
clustering models to describe reality accurately since our aim is to
compare the relative sensitivity of a two-point correlation function
analysis and the technique we have developed here. The validity of
this test is unlikely to depend strongly on the type of clustering.
We have explored two clustering scenarios:

1) Absorbers are clustered according to the gravitational
   quasi-equilibrium distribution (GQED) function \cite{Saslaw84}. We
   implement this scenario by following an approach first developed by
   \citeN{Neyman52} and described by \citeN{Sheth94}: we distribute
   clusters of absorbers randomly in a cosmological volume and draw
   the number of absorbers of a given cluster from the distribution
   \cite{Saslaw89}
\be
h(N) = \left \{
\begin{array}{ll}
b & N = 0 \\
\frac{N^{N-1}}{N!} (1-b) b^{N-1} {\rm e}^{-Nb} & N > 0.
\end{array} \right.
\ee
$b$ is the only parameter of the model and is defined as the ratio of
potential and kinetic energies of the cluster ($0 \le b \le 1$). It is
related to the two-point correlation function by \cite{Saslaw84}
\be 
b \equiv -\frac{W}{2K} = \frac{2 \pi G m^2 n}{3kT} \int_0^\infty 
\xi(r) dr,
\ee
where $T$ and $m$ are the temperature and mass of the cluster, $n$ is
the average number density and $k$ and $G$ have their usual meanings.
We choose $b = 0.3$ (\citeN{Sheth94} estimate for galaxies $b_0 \approx
0.75$) and members of a cluster have a velocity dispersion of
$500$~km~s$^{-1}$. We assume clusters to be spherical and distribute
absorbers within a cluster according to a King profile \cite{King66}.

2) Absorbers form `walls'. Considering the connection of the \lya\
   forest with galaxies at low redshift and the repeated findings of
   independent groups that galaxies form sheet- and wall-like
   structures (\citeNP{Broadhurst90}; \citeNP{Ettori97};
   \citeNP{Einasto97}; \citeNP{Connolly97}; \citeNP{DiNella96}) it is
   conceivable that such structures may also be found in the \lya\
   forest. In addition, at high redshift several hydrodynamic
   simulations have shown that the absorbing gas forms filaments,
   sheets and wall-like structures (\citeNP{Cen94};
   \citeNP{Miralda96}; \citeNP{CenSim97}; \citeNP{Hernquist96};
   \citeNP{Zhang95}; \citeNP{Petitjean95}; \citeNP{Muecket96};
   \citeNP{Riediger98}; \citeNP{Wadsley97}; \citeNP{Bond98}), although
   these structures are of a smaller scale than we are interested in.
   In any case, we have included this model where walls of absorber
   overdensities extend across several lines of sight in order to
   demonstrate the better sensitivity of our analysis compared to a
   conventional {\em cross}-correlation analysis of fitted absorption
   lines.

For both scenarios we have computed 100 sets of simulated spectra of a
close group of four QSOs using the parameters of S1.

\begin{figure}
\psfig{file=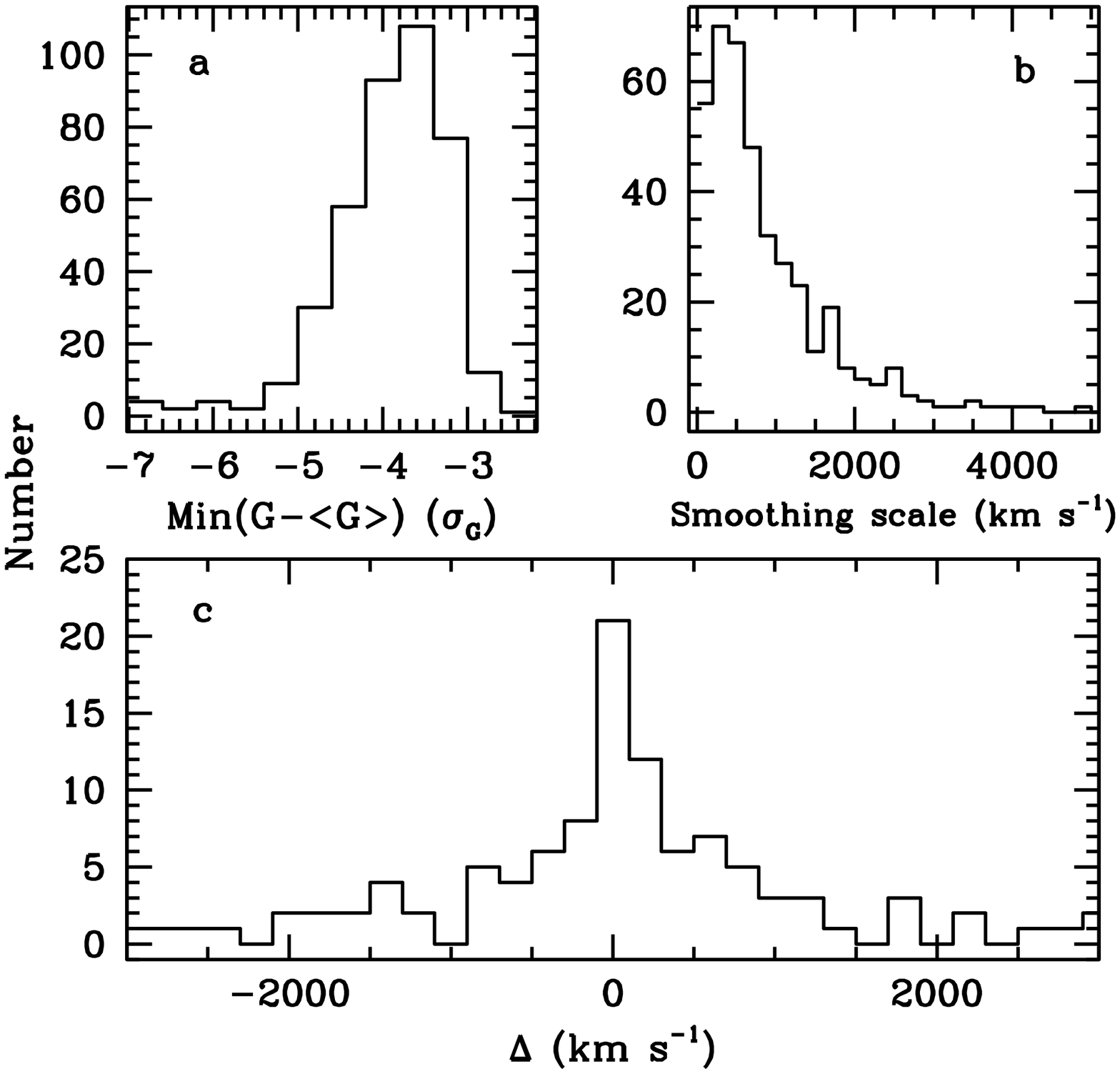,width=\textwidth,silent=}
\caption{Distributions of (a) values and (b) scales of minima detected
in reduced transmission triangles of spectra with GQED clustering.
(c) Distribution of differences between the positions of the
`strongest' clusters (see text) and the minima of (a).}
\label{comp_gqed1}
\end{figure}

Figure \mref{comp_gqed1} shows the result of our new analysis for the
case of GQED clustering. For all spectra we have computed their
transmission triangles, subtracted the mean given by equation
\mref{mG} and divided by the rms given by the square-root of equation
\mref{sigG}. We shall refer to the result as `reduced' transmission
triangles. In the reduced triangles all residual fluctuations are
given in terms of their statistical significance rather than their
absolute magnitude. In panel (a) of Figure \mref{comp_gqed1} we plot
the histogram of the minimum values (maximally significant overdense
absorption) measured in these reduced transmission triangles of the
individual spectra. The distribution peaks at $-3.6\sigma$ but in a
significant fraction of cases ($\sim 40$ per cent) we have a greater
than $4\sigma$ detection. Panels (b) and (c) show that these
detections are not spurious but actually arise from the clusters. In
panel (b) we plot the distribution of scales (FWHM of smoothing
Gaussian) at which the minima of panel (a) are detected. Clearly we
recover the correct velocity dispersion of the clusters. We loosely
define the `strongest' cluster in a spectrum as the cluster with the
highest total column density and plot in panel (c) the histogram of
differences in velocity space between the strongest clusters and the
detected minima, $\Delta$. Although there is clearly a peak at
$0$~km~s$^{-1}$ of the correct width, there are a large number of
cases where the detected minima do not coincide with the strongest
clusters. However, these mismatches do {\em not all} indicate spurious
detections. Rather, they are mostly due to our definition of the
strongest cluster, since it does not guarantee that the strongest
cluster will produce the maximum absorption.

We now compare the results above with a two-point correlation function
(tpcf) analysis. We compute both `real' and `observed' tpcfs from two
separate lists of absorption lines. A `real' list is derived from the
input line list used to create the spectrum by simply applying an
equivalent width detection threshold. To mimic blending due to
instrumental resolution we generate an `observed' line list from the
input line list by blending all lines that lie within one FWHM$_{{\rm
LSF}}$ of each other into a single line and imposing an equivalent
width detection limit. The position of the blended line is taken as
the equivalent width weighted average of its components. We estimate
the $3\sigma$ equivalent width detection limit in our simulated data
to be 0.26~\AA. The two-point correlation function is calculated as
\be 
\xi(\Delta v) = \frac{N_{\rm obs}(\Delta v)}{N_{\rm exp}
(\Delta v)} - 1, 
\ee 
where $N_{\rm obs}$ and $N_{\rm exp}$ are the observed and expected
number of pairs at separation $\Delta v$. We account for the evolution
of the mean line number density in the calculation of $N_{\rm
exp}$. The individual line correlation functions of a set of four
spectra are averaged to increase the signal to noise ratio.

\begin{figure}
\psfig{file=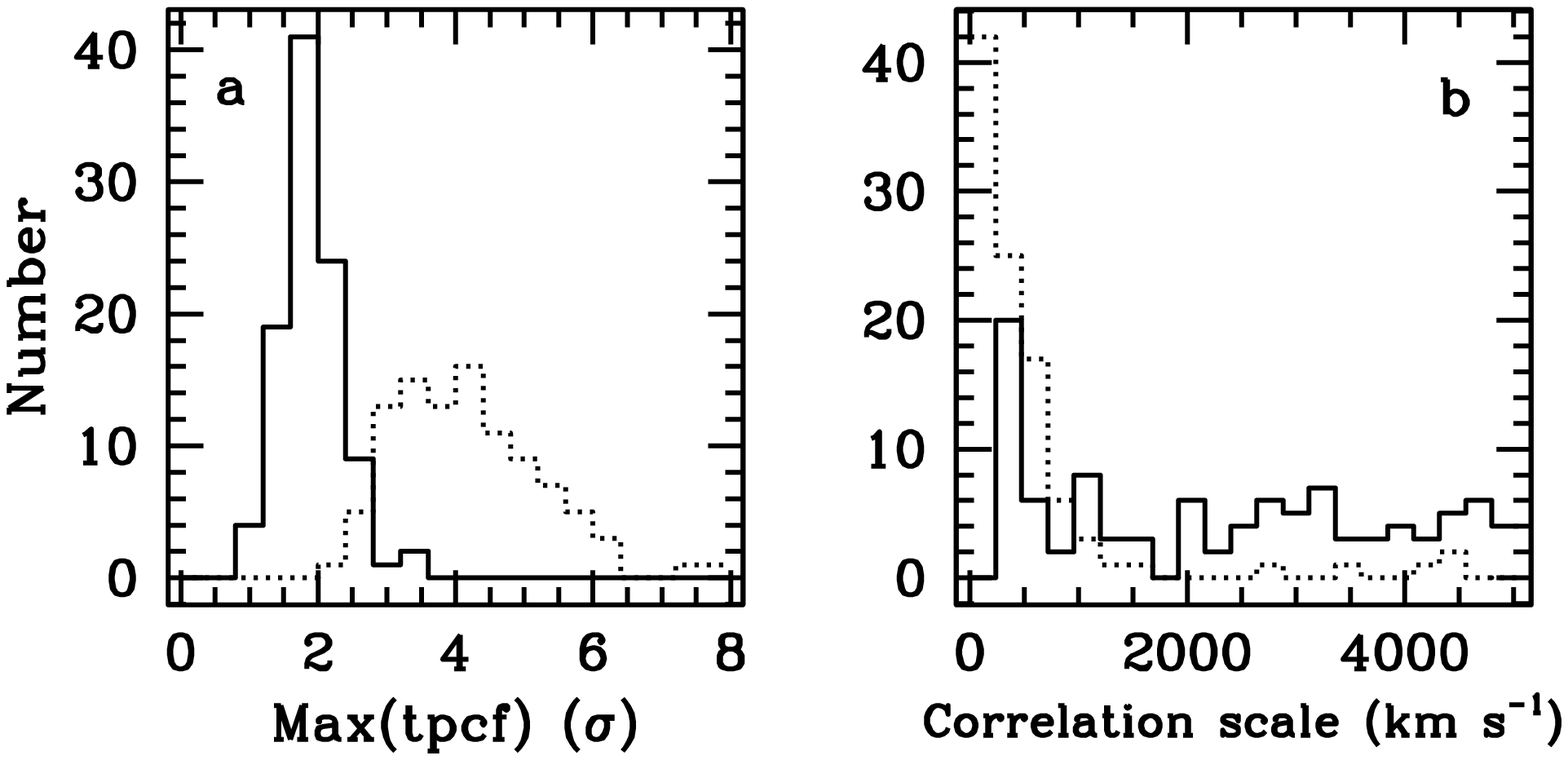,width=\textwidth,rheight=9cm,silent=}
\caption{Distributions of (a) values and (b) correlation scales of
averaged two-point correlation function maxima using `observed'
line lists (solid lines) and `real' line lists (dashed lines).}
\label{comp_gqed2}
\end{figure}
In panel (a) of Figure \mref{comp_gqed2} we show the distribution of
the maximally significant values detected in the averaged `observed'
(solid line) and `real' (dotted line) two-point correlation functions.
For an underlying clustered set of absorption lines, these
distributions will be slightly sensitive to the bin size chosen in
computing the tpcfs. To some extent this reflects one of the
difficulties with the tpcf; one must chose {\em a priori} a bin size,
without prior knowledge as to what an `optimal' size might be. In
practice, observers often chose the smallest convenient size which is
larger than the instrumental resolution. We have done similarly in
this experiment and have chosen $120$~km~s$^{-1}$.

The solid histogram in panel (a) peaks narrowly at $1.8\sigma$. Only 3
per cent of the detections are $> 3\sigma$. Panel (b) shows the
correlation scales at which the maxima are detected and we see that at
least 50 per cent of the detections are spurious. The dotted
histograms show the results for the `real' tpcfs: significant
detections (a) at the right scale (b). However, a comparison with
panel (a) of Figure \mref{comp_gqed1} shows that a tpcf analysis, even
with {\em infinite} resolution (but finite S/N) and a {\em perfect}
line fitting algorithm, does only marginally better in uncovering the
presence of clustering than our new analysis using intermediate
resolution.

\begin{figure}
\psfig{file=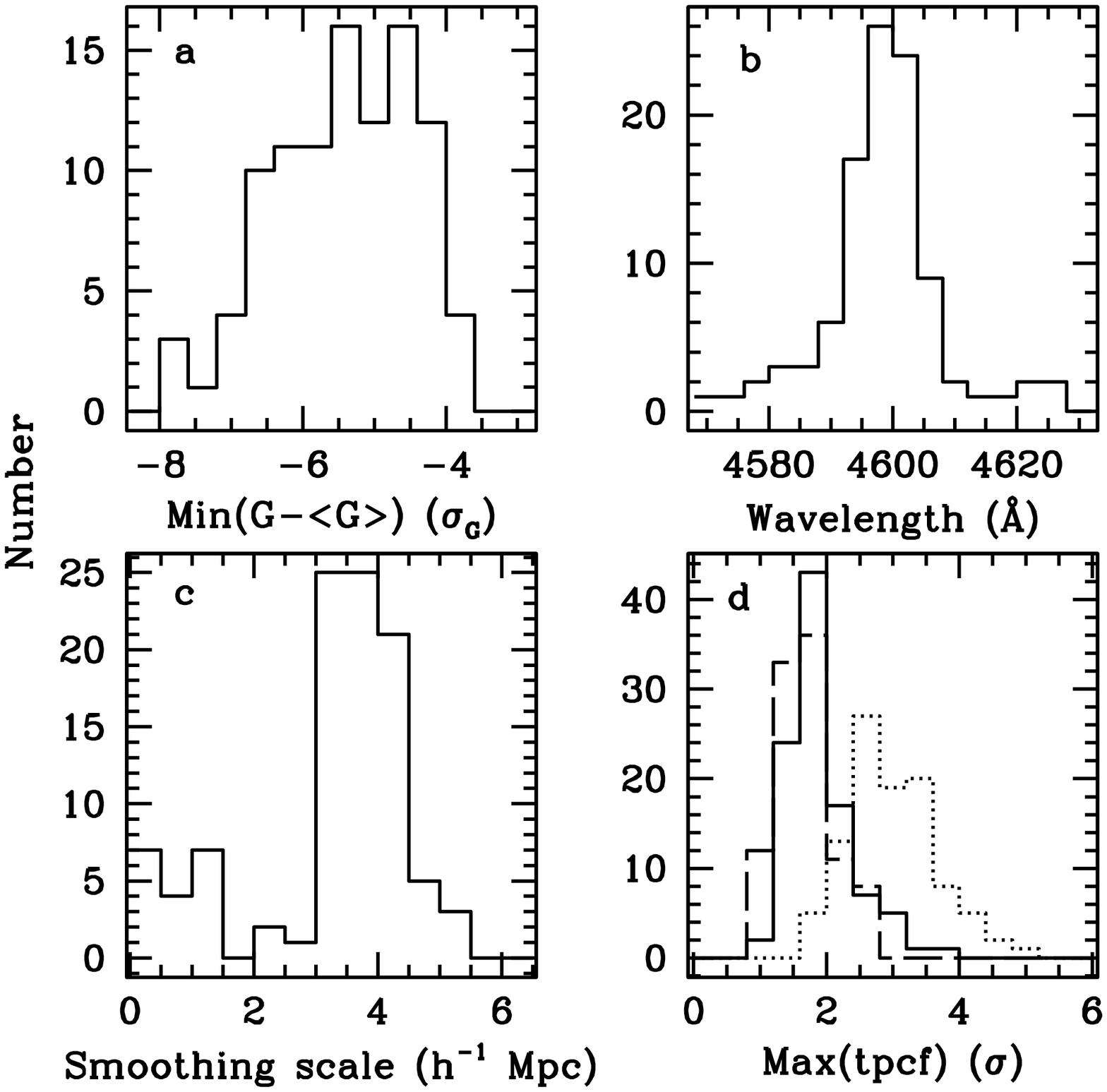,width=\textwidth,silent=}
\caption{Distributions of (a) values, (b) positions and (c) scales of
minima detected in reduced averaged transmission triangles, where
absorbers form a `wall' at 4600~\AA.  (d) Distribution of maxima of
averaged auto- (solid line) and cross-correlation (dashed line)
functions using `observed' and `real' (auto only, dotted line) line
lists.}
\label{comp_wall}
\end{figure}
Figure \mref{comp_wall} shows the results for the case of a `wall' of
absorbers which is simulated by multiplying the redshift distribution
of absorbers with a top hat function. The simulated wall is located at
$z = 2.78$, it is $5~h^{-1}$ Mpc thick and is overdense by a factor of
$\delta n = 2$. As described above we have averaged the individual
transmission triangles of each set of four spectra. The distributions
of the values, positions and scales of the minima detected in the
reduced averaged triangles are plotted in panels (a), (b) and (c)
respectively. {\em All} detections are above the $3\sigma$ level and
from panel (b) we see that all detections are due to the wall. Taking
the top hat shape of the wall into account, its thickness has
correctly been recovered in panel (c). Using the peaks of the three
distributions we calculate an overdensity of $2.6$ (see also Figure
\ref{sensi}). As in Figure \mref{comp_gqed2} we plot in panel (d) the
distribution of the maximum values detected in the averaged two-point
correlation functions using the `observed' (solid line) and the `real'
(dotted line) line lists. In addition, we performed a {\em
cross-}correlation analysis and show the result as the dashed
histogram. Both auto- and cross-correlations fail to deliver a
significant result. In fact, even with {\em infinite} resolution and a
{\em perfect} line fitting algorithm, the auto-tpcf analysis does a
worse job of uncovering the `wall' than our analysis using
intermediate resolution.

\begin{figure}
\psfig{file=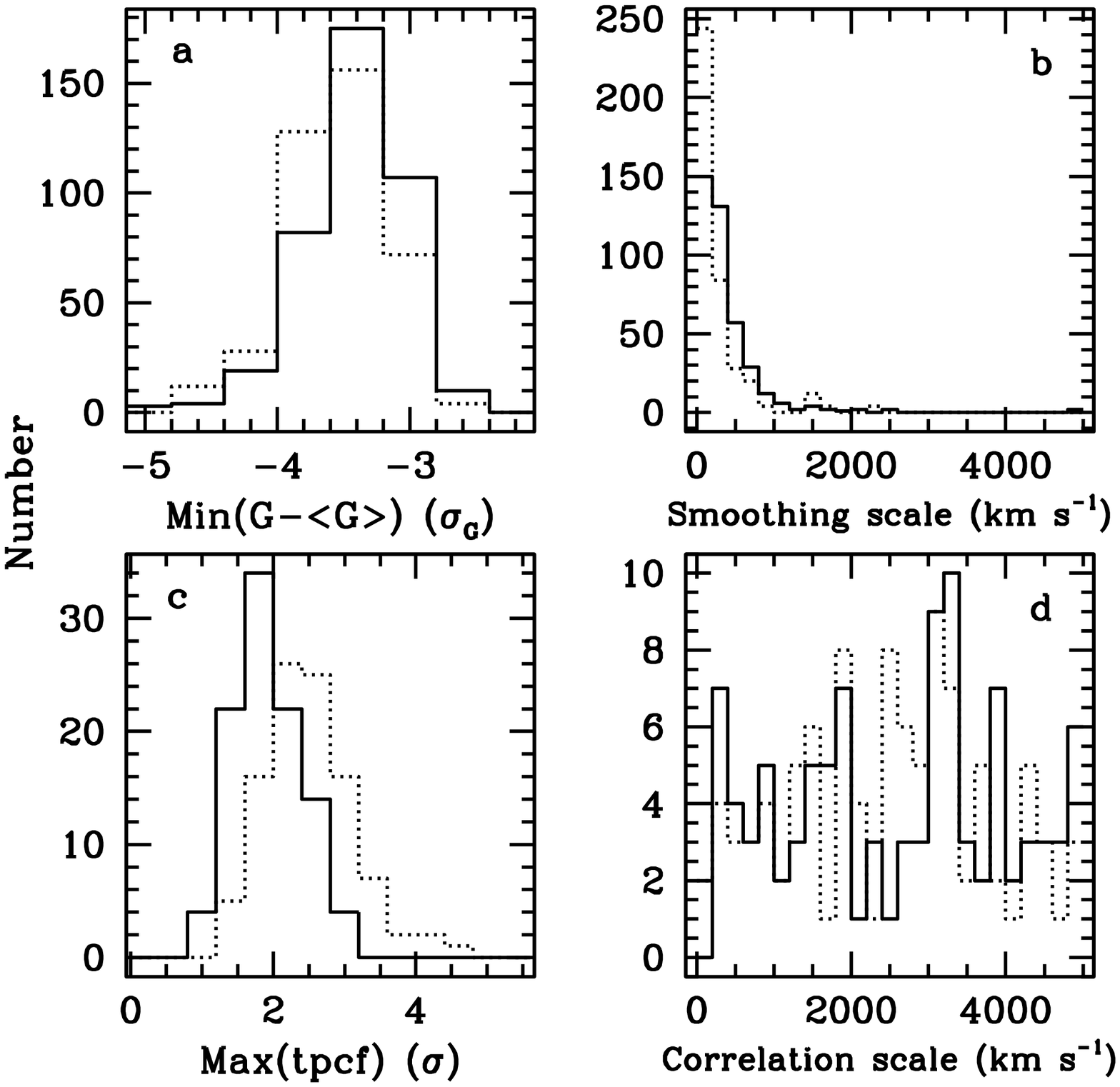,width=\textwidth,silent=}
\caption{Distributions of (a) values and (b) scales of minima detected
in reduced individual (solid lines) and averaged (dotted lines)
transmission triangles, where absorbers are distributed randomly. The
dotted histograms were renormalised. Distributions of (c) values and
(d) correlation scales of maxima of averaged two-point correlation
functions using `observed' (solid lines) and true (dotted lines) line
lists.}
\label{comp_rand}
\end{figure}

For both cases discussed above we have demonstrated that our new
analysis is substantially more sensitive to the presence of non-random
structure in the \lya\ forest than a traditional two-point correlation
function analysis when applied to intermediate resolution data. To
further illustrate this point we show in Figure \mref{comp_rand} the
same distributions as in Figures \mref{comp_gqed1}, \mref{comp_gqed2}
and \mref{comp_wall} for the case where absorbers are randomly
distributed. We note that the distributions of transmission minima in
Figures \mref{comp_gqed1} and \mref{comp_wall} differ substantially
from the one in Figure \mref{comp_rand}, whereas the distributions of
tpcf maxima are very similar. In panel (b) of Figure \mref{comp_rand}
we see the effect of the non-Gaussian statistics at small smoothing
scales as discussed above: the minimum value in a transmission
triangle is more likely to occur at small smoothing scales than at
large ones which is why the minima are not evenly distributed over all
scales as are the maxima of the tpcf.

\section{Conclusions} \label{conclusions}
In this paper we have developed a new technique to test for non-random
structure in the \lya\ forest. This new technique does not require
line fitting but is rather based on the statistics of the transmitted
flux. We have tested the relevant analytic calculations and
approximations against simulated data and have found excellent
agreement. We have argued that the accuracy of our method is limited
by the precision of the continuum fit and by the errors in the line
distribution parameters rather than by errors introduced by analytic
approximations. We have shown our new analysis to be substantially
more sensitive to non-randomness in intermediate resolution data than
a traditional two-point correlation function analysis. Finally, we
have presented evidence that, in the case of a coherent structure of
absorbers extending across several lines of sight, our analysis using
{\em intermediate} resolution data is at least comparable, if not
superior, in sensitivity to a tpcf analysis using {\em high}
resolution data. 

The next step is to apply our method to real data. In a forthcoming
paper we will present the results of our analysis of the spectra of a
close group of ten QSOs.

\label{lastpage}

\end{document}